\documentclass{webofc}
\usepackage[varg]{txfonts}   % Web of Conferences font
\usepackage{graphicx}
\woctitle{Physics at the Magnetospheric Boundary, Geneva, Switzerland (25-28 June, 2013)}
\def\beq#1{\begin{equation}\label{#1}}
\def\eeq{\end{equation}}
\def\beqa#1{\begin{eqnarray}\label{#1}}
\def\eeqa{\end{eqnarray}}

\def\Eq#1{(\ref{#1})}

\def\myfrac#1#2{\left(\frac{#1}{#2}\right)}
\def\comment#1{\relax}

\begin{document}
\title{Theory of wind accretion}

\author{N.I. Shakura\inst{1}\fnsep\thanks{\email{nikolai.shakura@gmail.com}} \and
        K.A. Postnov\inst{1}
%\fnsep\thanks{\email{Mail address for second author if necessary}} 
\and
        A.Yu. Kochetkova\inst{1}
%\fnsep\thanks{\email{Mail address for last author if necessary}} 
\and
	L. Hjalmarsdotter\inst{1}
        % etc.
}

\institute{Moscow M.V. Lomonosov State University, 
Sternberg Astronomical Institute, 13, Universitetskij pr., 119992 Moscow, Russia}

\abstract{%
A review of wind accretion in high-mass X-ray binaries is presented. We focus attention to different regimes of quasi-spherical accretion onto the neutron star: the supersonic (Bondi) accretion, which takes place when the captured matter cools down rapidly and falls supersonically toward NS magnetospghere, and subsonic (settling) accretion which occurs when plasma remains hot until it meets the magnetospheric boundary. Two regimes of accretion are separated by an X-ray luminosity of about $4\times10^{36}$~erg/s. In the subsonic case, which sets in at low luminosities, a hot quasi-spherical shell must be formed around the magnetosphere, and the actual accretion rate onto NS is determined by ability of the plasma to enter the magnetosphere due to Rayleigh-Taylor instability. We calculate the rate of plasma entry the magnetopshere and the angular momentum transfer in the shell due to turbulent viscosity appearing in the convective differentially rotating shell. 
We also discuss and calculate the structure of the magnetospheric boundary layer where 
the angular momentum between the rotating magnetosphere and the base of the 
differentially rotating quasi-spherical shell takes place. We show how observations of equilibrium X-ray pulsars Vela X-1 and GX 301-2 can be used to estimate dimensionless parameters of the subsonic settling accretion theory, and
obtain the width of the magnetospheric boundary layer for these pulsars.}
\maketitle
\section{Introduction}
\label{intro}
In close binary systems, 
there can be two different regimes of accretion onto the compact object -- disk accretion 
and quasi-spherical accretion. The disk accretion regime is usually realized 
when the optical star overfills its Roche lobe. Quasi-spherical accretion is 
most likely to occur in high-mass X-ray binaries (HMXB) 
when the optical star of early spectral class (O-B) does not fill its Roche lobe, 
but has a significant mass loss via stellar wind. We shall discuss the wind accretion regime,
in which a bow shock forms in the stellar wind near the compact star. The structure of the bow 
shock and accretion wake is quite complicated and is non-stationary (see e.g.
numerical simulations \cite{1988ApJ...335..862F}, \cite{1999A&A...346..861R}, \cite{2004A&A...419..335N}, 
among others). The characteristic distance at which the bow shock forms
is about the Bondi radius $R_B=2GM/(v_w^2+v_{orb}^2)$, where $v_w$ is the wind velocity 
(typically hundred-thousand km/s), $v_{orb}$ is the orbital velocity of the compact star.
Here we shall consider accretion onto magnetized neutron stars (NS) observed as X-ray pulsars. 
The rate of gravitational capture of mass from the wind (the Bondi mass accretion rate)
is $\dot M_B\simeq \rho_w R_B^2 v_w$.   

\begin{figure*}
\includegraphics[width=7cm]{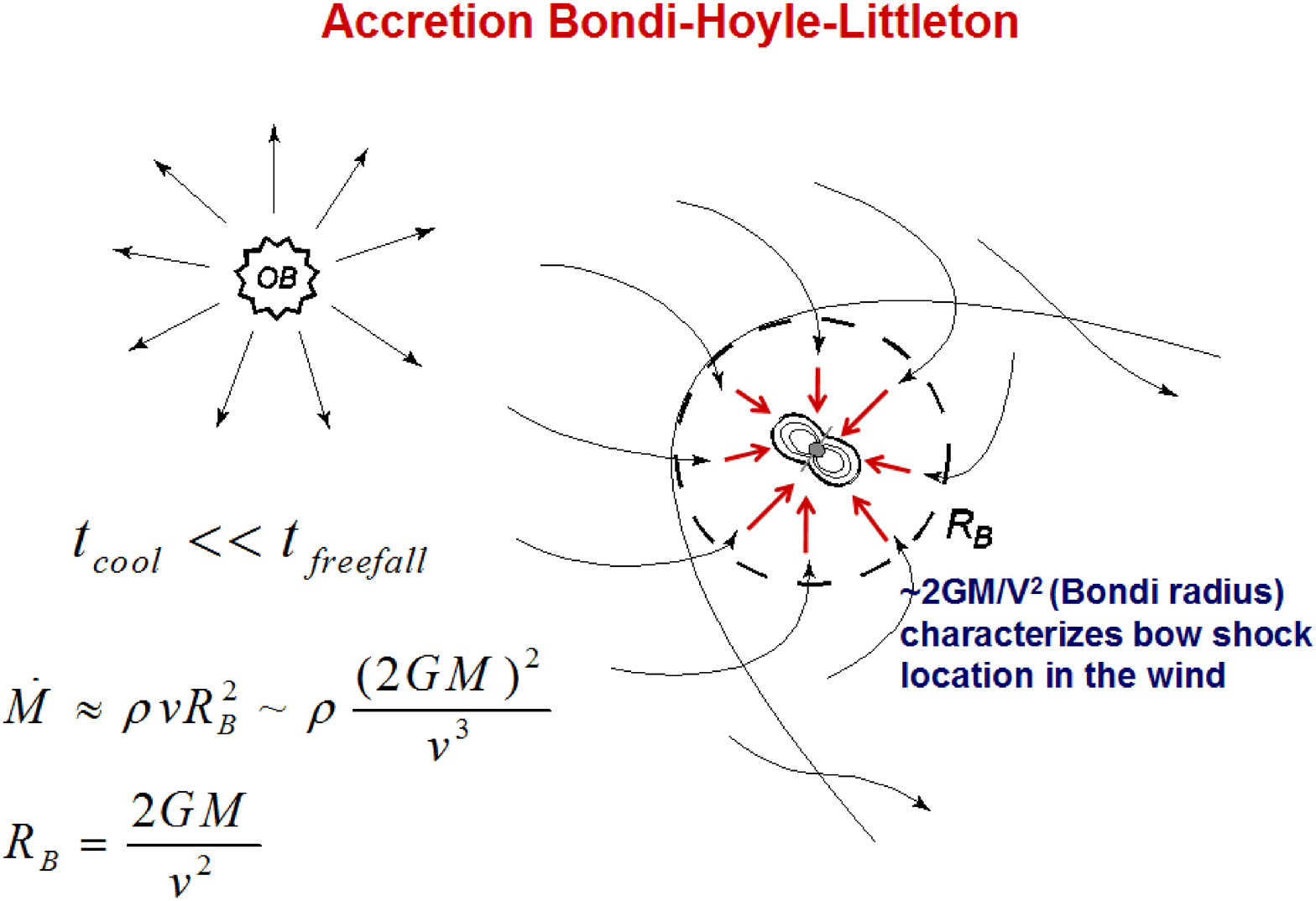}
\includegraphics[width=7cm]{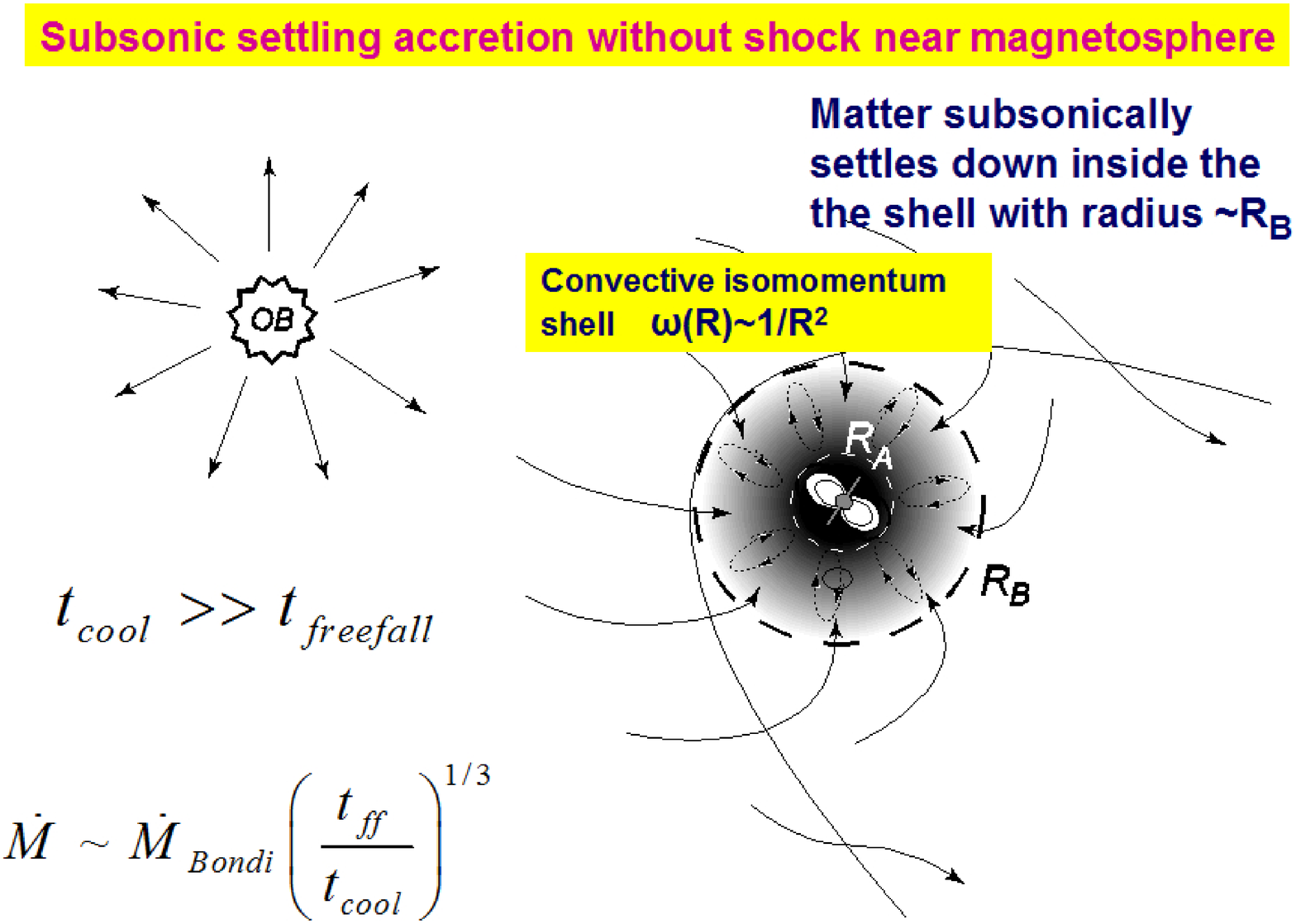}
\parbox[t]{0.47\textwidth}{\caption{Supersonic (Bondi-Hoyle-Littleton) accretion onto magnetized NS}\label{f:1}}
\hfill
%\caption
%{Best-fit orbital (top) and precessional (bottom) light curves of SS433 from {\it INTEGRAL} observations for model with $q=0.3$.}\label{q03}}
% \includegraphics[width=0.47\textwidth]{fig1E_obs_sets.ps}
%\includegraphics[width=0.45\textwidth]{ss433_orbit_hardness.ps}
%\hfill
%\parbox[t]{0.47\textwidth}{\caption{IBIS/ISGRI 18-40 and 40-60 keV orbital light curves with hardness
%ratio (bottom plot).}\label{f:horb}}
%\hfill
\parbox[t]{0.47\textwidth}
{\caption{Subsonic settling accretion onto magnetized NS}\label{f:2}}
%\caption
%{Best-fit orbital (top) and precessional (bottom) light curves of SS433 from {\it INTEGRAL} observations for model with $q=0.3$.}\label{q03}}
\end{figure*}

There can be two different cases of quasi-spherical accretion. The classical 
Bondi-Hoyle-Littleton 
accretion takes place when the shocked matter rapidly cools down (via 
Compton cooling), and the matter freely falls toward the NS magnetosphere.  
(see Fig. \ref{f:1}). 
A shock is formed at some distance above the magnetosphere. 
Above the magnetosphere, the shocked matter rapidly cools down 
and enters the magnetopshere via Rayleigh-Taylor instability \cite{1976ApJ...207..914A}.
The magnetospheric
boundary is characterized by the Alfv\'en radius $R_A$, which can be
calculated from the balance of the ram pressure of the 
infalling matter and the magnetic
field pressure at the boundary. 
The captured matter from the wind has a specific angular momentum 
$j_w\sim \omega_BR_B^2$ \cite{1975A&A....39..185I}. 
Depending on the sign of $j_w$ (prograde or retorgrade), the NS can spin-up 
or spin-down. This regime of quasi-sphericl accretion is realized in 
bright X-ray pulsars with $L_x>4\times 10^{36}$~erg/s \cite{2012MNRAS.420..216S}. 

If the shocked matter remains hot (when plasma cooling time 
is much longer than the free-fall time, $t_{cool}\gg t_{ff}$), 
a hot quasi-static shell forms above the magnetosphere. The subsonic 
(settling) accretion sets in (see Fig. \ref{f:2}).  
In this case, both spin-up or spin-down of the NS 
is possible, even if the sign of $j_w$ is positive (prograde). The shell mediates the
angular momentum transfer from the NS magnetosphere via viscous stresses
due to convection (see below). In this regime, the mean radial velocity 
of matter in the shell $u_r$ is smaller than the free-fall velocity $u_{ff}$: 
$u_r=f(u)u_{ff}$, $f(u)<1$, and is determined by the palsma cooling rate 
near the magnetosphere (due to Compton or radiative cooling): 
$f(u)\sim [t_{ff}(R_A)/t_{cool}(R_A)]^{1/3}$. Here the actual mass accretion rate
onto NS can be significantly smaller than the Bondi mass accretion rate, $\dot M=f(u) \dot M_B$.  
The settling accretion occurs 
at $L_x<4\times 10^{36}$~erg/s \cite{2012MNRAS.420..216S}. 

\section{Vertical structure of the subsonic shell}
\label{sec-1}
In the quasi-steady state, the structure of the shell is described by
hydrostatic equilibrium:
\beq{e1}
-\frac{1}{\rho}\frac{dP}{dR}-\frac{GM}{R^2}=0
\eeq 
with the adiabatic solution:
\beq{hse_sol}
\frac{{\cal R} T}{\mu_m} = \myfrac{\gamma-1}{\gamma}\frac{GM}{R}\,.
\eeq 
(Turbulence in the shell can play a certain role in hydrostatic equilibrium; this complicates
formulas but does not change the conclusions, see \cite{2012MNRAS.420..216S} for more detail). 
For $\gamma=5/3$ the density profile in the shell is:
\beq{rho(R)}
\rho(R)=\rho(R_A)\myfrac{R_A}{R}^{3/2}\,.
\eeq

The Alfv\'en surface is found from the balance of gas pressure and the magnetic field pressure
at the magnetopsheric boundary:
\beq{P(RA)}
P_g=\frac{B^2(R_A)}{8\pi}\,.
\eeq 
Taking into account the mass continuity equation in the settling regime, we obtain:
\beq{RA_def}
R_A=\left[\frac{4\gamma}{(\gamma-1)}f(u) K_2
\frac{\mu^2}{\dot M\sqrt{2GM}}\right]^{2/7}\,.
\eeq 
Here $K_2\approx 7.6$ is the numerical coefficient which takes into account the enhancement
of the magnetic field at the magnetopsheric boundary due to emerging currents \cite{1976ApJ...207..914A}. 

The plasma enters the magnetosphere of the slowly rotating neutron star due to 
the Rayleigh-Taylor instability. The boundary between the plasma and the magnetosphere is stable 
at high temperatures $T>T_{cr}$, but becomes unstable at $T<T_{cr}$, and remains in 
a neutral equilibrium at $T=T_{cr}$ \cite{1977ApJ...215..897E}. The critical temperature is:
\beq{Tcr}
{\cal R}T_{cr}=\frac{1}{2}\frac{\cos\chi}{\kappa R_A}\frac{\mu_mGM}{R_A}\,.
\eeq 
Here $\kappa$ is the local curvature of the magnetosphere, $\chi$ is the angle 
the outer normal makes with the radius-vector at a given point. The effective gravity acceleration can be written as
\beq{g_eff}
g_{eff}=\frac{GM}{R_A^2}\cos\chi\left(1-\frac{T}{T_{cr}}\right)\,.
\eeq 
The temperature in the quasi-static shell is given by \Eq{hse_sol}, and 
the condition for the magnetosphere instability can thus be rewritten as:
\beq{m_inst}
\frac{T}{T_{cr}}=\frac{2(\gamma-1)}
{\gamma}\frac{\kappa R_A}{\cos\chi}<1\,.
\eeq 
Let us consider the development of the interchange instability when cooling 
(predominantly Compton cooling) is present. The temperature changes as 
\cite{kompaneets56}, \cite{1965PhFl....8.2112W}
\beq{dTdt}
\frac{dT}{dt}=-\frac{T-T_x}{t_C}\,,
\eeq 
where the Compton cooling time is
\beq{t_comp}
t_{C}=\frac{3}{2\mu_m}\frac{\pi R_A^2 m_e c^2}{\sigma_T L_x}%\frac{1}{1-T_x/T_e}
\approx 10.6  [\hbox{s}] R_{9}^2 \dot M_{16}^{-1}\,.%\frac{1}{1-T_x/T_e}\,.
\eeq
Here $m_e$ is the electron mass, $\sigma_T$ is the Thomson cross section, $L_x=0.1 \dot M c^2$ is the X-ray luminosity, $T$ is the electron temperature (which is equal to the ion temperature since the timescale of electron-ion energy exchange here is the shortest possible), $T_x$ is the X-ray temperature and
$\mu_m=0.6$ is the molecular weight. The photon temperature is $T_x=(1/4) T_{cut}$ for a bremsstrahlung spectrum with an exponential cut-off at $T_{cut}$, typically $T_x=3-5$~keV. 
The solution of equation \Eq{dTdt} reads:
\beq{}
T=T_x+(T_{cr}-T_x)e^{-t/t_C}\,.
\eeq 
We note that $T_{cr}\sim 30\,\hbox{keV}\gg T_x\sim 3$~keV. It is seen that for $t\approx 2t_C$ the temperature decreases to $T_x$. In the linear approximation the temperature changes as:
\beq{tlin}
T\approx T_{cr}(1-t/t_C)\,.
\eeq 
Plugging this expression into \Eq{g_eff}, we find that the effective gravity acceleration increases linearly with time as:
\beq{}
g_{eff}\approx \frac{GM}{R_A^2}\frac{t}{t_C}\cos\chi \,.
\eeq 
Correspondingly, the velocity of matter due to the instability growth increases with time as:
\beq{u}
u_r=\int\limits_0^{t} g_{eff} dt=\frac{GM}{R_A^2}\frac{t^2}{2t_C}\cos\chi \,.
\eeq 
Let us introduce the mean rate of the instability growth
\beq{}
<u_i>=\frac{\int u dt}{t}=\frac{1}{6}\frac{GM}{R_A^2}\frac{t^2}{t_C}=
\frac{1}{6}\frac{GM}{R_A^2t_C}\myfrac{\zeta R_A}{<u_i>}^2\,.
\eeq
Here $\zeta\lesssim 1$ and $\zeta R_A$ is the characteristic scale of the
instability that grows with the rate $<u_i>$.
So for the mean rate of the instability growth in the linear stage we find
\beq{ui}
<u_i>=\myfrac{\zeta^2GM}{6t_C}^{1/3}=\frac{\zeta^{2/3}}{12^{1/3}}\sqrt{\frac{2GM}{R_A}}
\myfrac{t_{ff}}{t_C}^{1/3}\,.
\eeq
Here we have introduced the free-fall time as
\beq{}
t_{ff}=\frac{R_A^{3/2}}{\sqrt{2GM}}\,.
\eeq
Therefore, the factor $f(u)$ becomes:
\beq{fu1}
f(u)=\frac{<u_i>}{u_{ff}(R_A)}\,.
\eeq
Substituting \Eq{ui} and \Eq{fu1} into \Eq{RA_def}, we find for the Alfven radius in 
this regime:
\beq{RA}
R_A\approx 1.37\times 10^9[\hbox{cm}]
\left(\zeta\frac{\mu_{30}^3}{\dot M_{16}}\right)^{2/11}\,.
\eeq
(We stress the difference of the obtained expression for the Alfven radius with the 
standard one, $R_A\sim \mu^{4/7}/\dot M^{-2/7}$). 
Plugging \Eq{RA} into \Eq{fu1}, we obtain explicit expression for $f(u)$:
\beq{fu}
f(u)\approx 0.22
\zeta^{7/11}\dot M_{16}^{4/11}\mu_{30}^{-1/11}\,.
\eeq

In principle, for moderately rotating X-ray puslars 
plasma can entry a rotating magnetosphere via 
Kelvin-Helmholtz instability \cite{1983ApJ...266..175B}:
\beq{}
u_{KH}\simeq 0.1 \gamma_{KH}/k=0.1\left\{
\begin{array}{cc}
u_{\phi}\frac{\sqrt{\rho_i/\rho_e}}{1+\rho_i/\rho_e},&\myfrac{u_A}{c}^2<\frac{\rho_i}{\rho_e}\\
u_{\phi}\frac{u_A}{c},&\myfrac{u_A}{c}^2>\frac{\rho_i}{\rho_e}\\
\end{array}
\right.
\eeq
Here $\rho_e$ is the external density (above $R_A$), 
$\rho_i$ is the internal density (below $R_A$), $u_A=B/\sqrt{4\pi\rho_e}$ is the Alfv\'en 
velocity. In the settling accretion regime $u_{RT}\approx 0.31 u_{ff}$ (see above). 
Clearly, for slowly rotating pulsars with $u_{ff}\gg u_\phi=\omega^*R_A$ the plasma
entry rate $u_{KH}\ll u_{RT}$ and Kelvin-Helmholtz instability is ineffective. 

\section{Spin-up/spin-down during settling accretion}

In our problem there are three characteristic angular frequencies: 
the angular orbital frequency $\omega_b=2\pi/P_b$, 
which characterizes the specific angular momentum of captured matter, 
the angular frequency of matter near the magnetosphere, $\omega_m(R_A)$, and
the angular frequency of magnetosphere $\omega^*=2\pi/P^*$ which coincides with the NS angular 
rotation frequency. If $\omega_m(R_A)-\omega^*\ne 0$, an effective exchange 
of angular momnetum between the magnetosphere and the quasi-spherical shell occurs. 
As shown in Appendices in \cite{2012MNRAS.420..216S}, \cite{2013arXiv1302.0500S},
the rotational law in the shell with settling accretion can be represented in 
a power-law from $\omega(R)\sim 1/R^n$, with $0\le n\le 2$ depending on the
treatment of viscous stresses $W_{R\phi}$ in the shell. In the most likely case where 
anisotropic turbulence appears due to near-sonic convection, $n\approx 2$, i.e.
iso-angular-momentum rotaional law sets in.    

Magnetospheric torques applied to the NS can be written as 
\beq{torquem}
I\dot \omega^*=\int\frac{B_tB_p}{4\pi}\varpi dS 
\eeq
where $I$ is the neutron star's moment of inertia, 
$\varpi$ is the distance from the rotational axis. There can be different
cases of the coupling of
accreting plasma with rotating NS magnetosphere. 

\textbf{a) Strong coupling}.
This regime is likely to be realized when accreting palsma is magnetized. 
Powerful large-scale convective motions may lead to turbulent magnetic field diffusion accompanied by magnetic field dissipation. 
This process is characterized by the turbulent magnetic field diffusion coefficient 
$\eta_t$.  In this case 
the toroidal magnetic field (see e.g. \cite{1995MNRAS.275..244L} and references therein) is:
\beq{bt}
B_t=\frac{R^2}{\eta_t}(\omega_m-\omega^*)B_p\,.
\eeq
The turbulent magnetic diffusion coefficient is related to the kinematic turbulent viscosity as
$\eta_t\simeq \nu_t$. The latter can be written as:
\beq{nut}
\nu_t=<u_tl_t>\,.
\eeq  
According to the phenomenological Prandtl law, the average characteristics of 
a turbulent flow (the velocity $u_t$, the characteristic scale of turbulence $l_t$ 
and the shear $\omega_m-\omega^*$) are related as:
\beq{Prandtl}
u_t\simeq l_t |\omega_m-\omega^*|\,.
\eeq
In the case of turbulent magnetic diffusion, there is 
no narrow boundary layer, and the exchange of angular momentum between the magnetosphere and the shell 
occurs on the scale $\sim R_A$, i.e. $l_t\simeq R_A$, 
which determines the turn-over velocity of the largest turbulence eddies. At smaller scales a turbulent cascade develops. Substituting this scale into equations \Eq{bt}-\Eq{Prandtl} above, we find that in the strong coupling regime $B_t\simeq B_p$.
Then from \Eq{torquem} we find
\beq{17a}
I\dot \omega^*=\int\frac{B_tB_p}{4\pi}\varpi dS = 
\pm \tilde K(\theta)K_2\frac{\mu^2}{R_A^3}
\eeq
where $\tilde K(\theta)$ is a numerical coefficient depending
on the angle between the rotational and magnetic dipole axes. 

\textbf{b) Moderate coupling and structure of magnetospheric boundary layer}. 
Mechanical torque acting on the magnetosphere from the base 
of the shell due to turbulent stresses $W_{R\phi}$ is:
\beq{torquet}
\int W_{R\phi} \varpi dS\,,
\eeq
where the viscous turbulent stresses can be written as 
(see the Appendices in \cite{2012MNRAS.420..216S}, \cite{2013arXiv1302.0500S}
for more detail)
\beq{wrfi}
W_{R\phi}=\rho \nu_t R \frac{\partial \omega}{\partial R}\,.
\eeq
The turbulent viscosity coefficient in the boundary layer is 
$
\nu_t=\langle u_t l_t\rangle\,.
$ 
%with $l_t\sim \zeta' R_A$, the coefficient $\zeta'<1$ characterizes
%the size of the boundary layer. 
According to Prandtl's law, we write
\footnote{In paper \cite{2013arXiv1302.0500S} we did not use the empirical Prandtl rule, but
final results turn out to be insensitive to the turbulent viscosity treatment, cf. Table 1 here 
and Table in \cite{2013arXiv1302.0500S}.}
$
u_t=l_tR d\omega/dR
$
and viscous stresses in the form
\beq{wrfi2}
W_{R\phi}=\rho l_t^2 (R d\omega/dR)^2\,.
\eeq
Next we assume the characteristic turbulence length around 
non-spherically shaped magnetosphere 
$l_t=\kappa' R_A$, where $\kappa'\equiv R_A/<R_{curv}>-1\simeq 0.3$ 
(here $<R_{curv}>$ is the mean curvature radius of the magnetopshere).
For a rotating sphere we would have $\kappa'=0$, and $l_t=\kappa''(R-R_A)$ 
with $\kappa''$ being some constant \cite{Loiz}.   
In the magnetospheric boundary layer we shall omit
the advective term $\sim \dot M\omega R^2$, which can be neglected 
for low radial velocities $u_R\ll (B_t/B_p)u_A$. Then, integrating \Eq{wrfi2} we find 
\beq{lnprof}
R_A(\omega_m-\omega^*)=\frac{1}{\kappa'}\sqrt{\frac{W_{R\phi}}{\rho}}\ln\myfrac{R}{R_A}\,.
\eeq
Therefore, in our problem we recover the characteristic logarithmic 
profile of  velocity in the boundary layer, which is typical for boundary layers in general
\cite{Loiz}.  
At small distances from the Alfven surface $R=R_A+\Delta$, $\Delta\ll R_A$, 
the logarthimic profile of the angular velocity becomes linear:
\beq{}
R_A(\omega_m-\omega^*)=\frac{1}{\kappa'}\sqrt{\frac{W_{R\phi}}{\rho}}\frac{\Delta}{R_A}\,.
\eeq
At $R\gg R_A$ the angular velocity profile becomes iso-angular-momentum (see above). 
On the other hand, at the magnetospheric boundary the viscous torque
can be equated to magnetic torques $W_{R\phi}|_{R_A}=[B_tB_p/(4\pi)]|_{R_A}$. Therefore, 
in the bottom of the boundary layer with linear velocity profile, 
where we shall assume the strongest angular momentum transfer 
between the magnetosphere and the shell to occur, we obtain:
\beq{BtBp}
\left|\frac{B_t}{B_p}\right|=\frac{\gamma \kappa'^2}{2u_s^2}\myfrac{\omega_m-\omega^*}{\zeta'}^2R_A^2
\eeq
where $\zeta'\equiv \Delta/R_A < 1$ is the dimensionless thickness
of the boundary layer. 
%(To obtain this formula, we 
%have used the pressure balance at the magnetospheric boundary \Eq{P(RA)} 
%and the expression for the temperature \Eq{hse_sol}). 
After introducing
the Keplerian velocity $\omega_K(R_A)$ and integrating
over the magnetospheric boundary, we finally find:
\beq{sd1}
I\dot \omega^*=\tilde K\frac{2\gamma \kappa'^2}{\gamma-1}K_2\frac{\mu^2}{R_A^3}
\myfrac{\omega_m-\omega^*}{\omega_K(R_A)}\left|\frac{\omega_m-\omega^*}{\zeta'^2\omega_K(R_A)}\right|
\eeq
 where the geometrical factors arising from the integration of 
\Eq{torquem} are included in the coefficient $\tilde K\sim 1$. 
It is also convenient to introduce coupling coefficients $K_1=2\gamma \kappa^2/(\gamma-1)$
(which for $\gamma=5/3$ is $K_1=5\kappa'^2\sim 1$) and 
\beq{K3}
K_3\equiv \left|\frac{\omega_m-\omega^*}{\zeta'^2\omega_K(R_A)}\right|\,.
\eeq
Using hydrodynamic similarity principles, we shall assume that 
$K_3=const$, and in real pulsars we find $K_3\simeq 10$ (see below), suggesting 
the reasonable 
width of the boundary layer $\zeta'^2\sim 0.1(\omega_m-\omega^*)/\omega_K(R_A)$. 

Using the definition of the Alfv\'en radius $R_A$ \Eq{RA_def} and the expression for the Keplerian frequency $\omega_K$, we can write \Eq{sd1} in the form
\beq{sd_om}
I\dot \omega^*=Z \dot M R_A^2(\omega_m-\omega^*).
\eeq
Here the dimensionless coefficient $Z$ is 
\beq{Zdef}
Z=\frac{\tilde KK_1K_3}{f(u)}\frac{\sqrt{2}(\gamma-1)}{4\gamma}\,.
\eeq
Substituting in this formula $\gamma=5/3$ and the expression  \Eq{fu1}, we find 
\beq{Znum}
Z\approx 0.64 \tilde K K_1 K_3\zeta^{-7/11}\dot M_{16}^{-4/11}\mu_{30}^{1/11}.
\eeq  

Taking into account that the matter that falls onto the neutron star adds the angular momentum
$z\dot M R_A^2\omega^*$, we ultimately get 
\beq{sd_eq}
I\dot \omega^*=Z \dot M R_A^2(\omega_m-\omega^*)+z \dot M R_A^2\omega^*\,.
\eeq
Here $0<z<1$ is the numerical coefficient which is $\sim 2/3$ if matter
enters across the magnetospheric surface with equal probability. 
Substituting $\omega_m(R_A)=\omega_B(R_B/R_A)^2$ for iso-angular-momentum shell, we can rewrite
the above equation in the form
\beq{sd_eq1}
I\dot \omega^*= Z\dot M \omega_B R_B^2-Z(1-z/Z)\dot M R_A^2\omega^*\,,
\eeq 
or in the form explicitely showing spin-up and spin-down torques:
\beq{sd_eq2}
I\dot \omega^*=A\dot M^{\frac{7}{11}} - B\dot M^{3/11}\,.
\eeq

 \begin{figure*}
\includegraphics[width=0.5\textwidth]{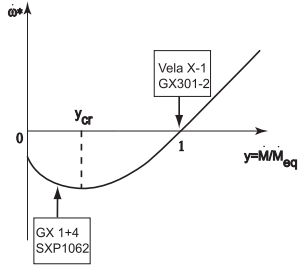}
\caption{An illustration of the dependence of $\dot\omega^*$ on the dimensionless accretion rate $y$. 
In fact as $y\to 0$, $\dot \omega^*$ approaches some negative value since the neutron star 
enters the propeller regime at small accretion rates. The figure shows the position in the diagram for equilibrium pulsars with $y\sim 1$ and for non-equilibrium pulsars at steady spin-down with $y<y_{cr}$}
\label{f:y}
 \end{figure*}

For a characteristic value of the accretion rate $\dot M_{16}\equiv \dot M/10^{16}$~g/s, the coefficients (not dependent on the accretion rate) will be equal to (in CGS units):
\beq{A(Z)}
A\approx 7.46\times 10^{31} \tilde K K_1 K_3\zeta^{-\frac{7}{11}}
\mu_{30}^{\frac{1}{11}}\myfrac{v_8}{\sqrt{\delta}}^{-4}\myfrac{P_b}{10\hbox{d}}^{-1}
\eeq
\beq{B}
B\approx 6.98 \times 10^{32}(1-z/Z) \tilde K K_1 K_3 \zeta^{-3/11}\mu_{30}^{{13}/{11}}
\myfrac{P^*}{100\hbox{s}}^{-1}
\eeq
The dimensionless factor $\delta<1$ takes into account the actual location of the gravitational capture radius.

\section{Equilibrium pulsars}

\begin{figure*}
\includegraphics[width=0.45\textwidth]{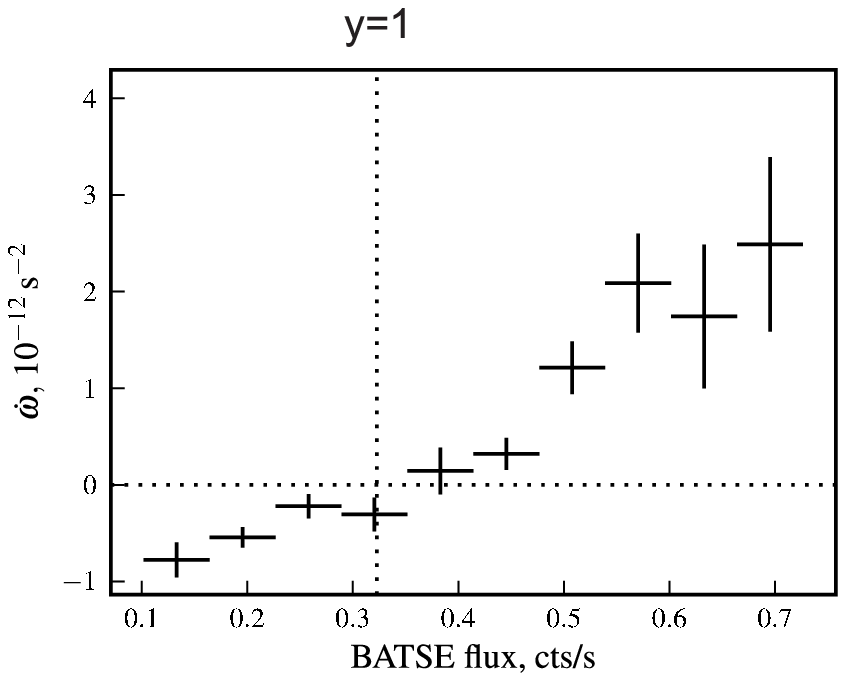}
\includegraphics[width=0.45\textwidth]{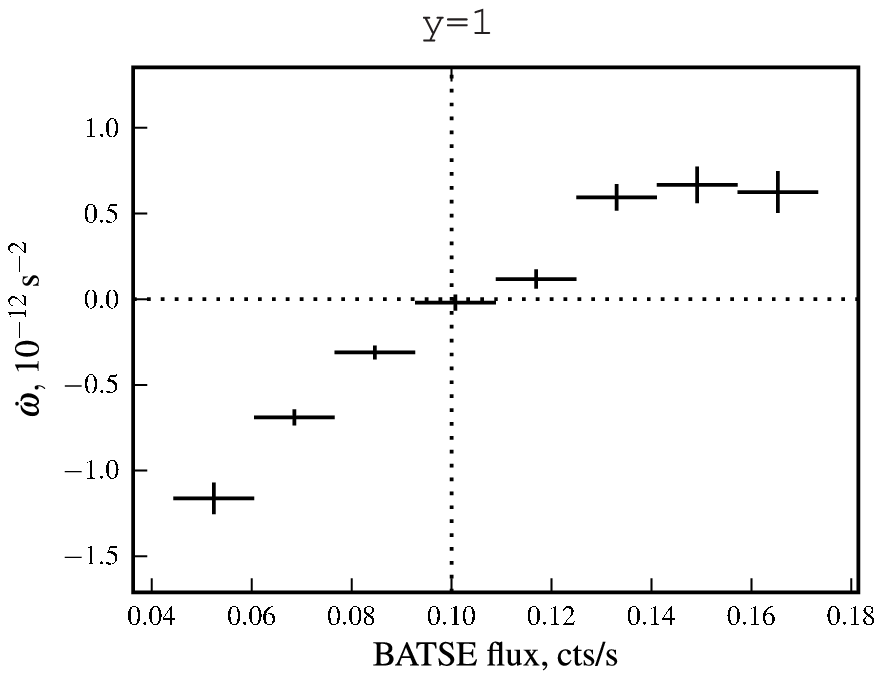}
\parbox[t]{0.47\textwidth}{\caption{Torque-luminosity correlation in GX 301-2, $\dot\omega^*$ as a function of BATSE data (20-40 keV pulsed flux) near the equilibrium frequency \cite{Dor10}. 
The assumed X-ray flux at equilibrium (in terms of
the dimensionless parameter $y$) is also shown by the vertical dotted line.}\label{f:gx}}
\hfill
%\caption
%{Best-fit orbital (top) and precessional (bottom) light curves of SS433 from {\it INTEGRAL} observations for model with $q=0.3$.}\label{q03}}
% \includegraphics[width=0.47\textwidth]{fig1E_obs_sets.ps}
%\includegraphics[width=0.45\textwidth]{ss433_orbit_hardness.ps}
%\hfill
%\parbox[t]{0.47\textwidth}{\caption{IBIS/ISGRI 18-40 and 40-60 keV orbital light curves with hardness
%ratio (bottom plot).}\label{f:horb}}
%\hfill
\parbox[t]{0.47\textwidth}
{\caption{The same as in Fig. \ref{f:gx} for Vela X-1 (V.Doroshenko, PhD Thesis, 2010, IAAT)}\label{f:vela}}
%\caption
%{Best-fit orbital (top) and precessional (bottom) light curves of SS433 from {\it INTEGRAL} observations for model with $q=0.3$.}\label{q03}}
\end{figure*}

For equilibrium pulsars we set 
$\dot \omega^*=0$ and from Equation \Eq{sd_eq} we get
\beq{Zeq}
Z_{eq}(\omega_m-\omega^*)+z\omega^*=0\,.
\eeq 
Close to equilibrium we may vary \Eq{sd_eq} with respect to $\dot M$. It is convenient to introduce the dimensionless parameter $y\equiv \dot M/\dot M_{eq}$, so that close to equilibrium $y=1$. Clearly, 
$\dot M_{eq}$ represents the accretion rate at which $\dot\omega^*=0$:
\beq{dotmeq}
\dot M_{eq}=\myfrac{B}{A}^{11/4}\,.
\eeq

Close to equilibrium we may vary \Eq{sd_eq} with respect to $\dot M$.
Variations in $\delta \dot M$ may in general be caused by changes in density $\delta \rho$ as well as in velocity of the stellar wind $\delta v$ (and thus the Bondi radius). For density variations only we find
\beq{Zeqrho}
Z_{eq,\rho}=\frac{I\frac{\partial \dot\omega^*}{\partial \dot M}|_{eq}}{\frac{4}{11}\omega^*R_A^2}\approx \ 2.52\myfrac{\frac{\partial \dot\omega^*}{\partial y}|_{y=1}}{10^{-12}}\myfrac{P^*}{100s}
\zeta^{-4/11}\dot M_{16}^{-7/11}\mu_{30}^{-12/11}\,.
\eeq
The equilibrium period of an X-ray pulsar with known NS magnetic field is:
\beq{Peq}
P_{eq}\approx 940 [\hbox{s}](1-z/Z_{eq})\zeta^{4/11}\mu_{30,eq}^{12/11}\myfrac{P_b}{10\hbox{d}}
\dot M_{16}^{-4/11}\myfrac{v_8}{\sqrt{\delta}}^{4}\,.
\eeq
Because of the strong dependence of the equilibrium period on wind velocity, 
for pulsars with independently known magnetic fields $\mu$  
it is more convenient to estimate the wind velocity, assuming $P*=P^*_{eq}$:
\beq{e:v8min}
\frac{v_8}{\sqrt{\delta}}\approx 0.57 (1-z/Z_{eq})^{-1/4}
\zeta^{-1/11} \dot M_{16}^{1/11}\mu_{30,eq}^{-3/11}
\myfrac{P_*/100 \hbox{s}}{P_b/10 \hbox{d}}^{1/4}\,.
\eeq
%
%Near equilibrium we get from \Eq{sd_eq2}:
%\beq{mu_eq}
%\mu_{30}^{(eq)}\approx
%0.13 (1-z/Z)^{-\frac{11}{12}} \zeta^{-1/3}
%\myfrac{\sqrt{\delta}}{v_8}^\frac{11}{3} \dot M_{16}^{1/3}
%\myfrac{P_*/100 \hbox{s}}{P_b/10 \hbox{d}}^\frac{11}{12}
%\,.
%\eeq
If $(\partial \dot\omega/\partial y)_{eq}$ is also measured, then  
equating $Z_{eq}$ to the rhs of \Eq{Znum} we find the value of the magnetic moment of the neutron star only from the pulsar equilibrium period and the derivative $(\partial \dot\omega/\partial y)_{eq}$:
\beq{mueqnew}
\mu_{30,eq}^{13/11}\approx 3.9 \myfrac{\frac{\partial \dot\omega^*}{\partial y}|_{y=1}}{10^{-12}\hbox{rad/s}^2}\myfrac{P^*}{100s}
(\tilde K K_1 K_3)^{-1}\zeta^{3/11}\dot M_{16}^{-3/11}\,.
\eeq 
If $\mu$ is independently measured, \Eq{mueqnew} allows us to 
determine the dimensionless complex of coefficients of the theory $\tilde K K_1 K_3\zeta^{-3/11}$:
 \beq{k1k3zeta}
\tilde K K_1 K_3\zeta^{-3/11}\approx 3.9 \myfrac{\frac{\partial \dot\omega^*}{\partial y}|_{y=1}}{10^{-12}\hbox{rad/s}^2}\myfrac{P^*}{100s}\dot M_{16}^{-3/11}\mu_{30}^{-13/11}\,.
\eeq

\begin{table*}
\label{t:tab}
 \centering
 \caption{Parameters for the equilibrium X-ray pulsars.} 
%References for the observed pulsar and orbital parameters are given in the text as well as values for the wind velocities from measurements of the optical components. The parameters $Z$, $K_1/\zeta$ and $f(u)$ were derived in Sections \ref{s:f(u)} and \ref{s:angmom}.
%Numerical estimates are given assuming iso-angular-momentum rotation in the shell ($n=2$), moderate coupling between the plasma and the magnetic field $\delta=1, \zeta=1$, $\tilde\omega=1$, $K_0=1$, $\gamma=5/3$ without turbulence ($m_t=0$, $K_t=1$).} 
 $$
\begin{array}{lcc}
\hline
\hbox{Pulsar }&\multicolumn{2}{c}{\hbox{Equilibrium pulsars }} \\
%\multicolumn{3}{c}{\hbox{non-equilibrium pulsars }}\\
\hline
& {\rm GX 301-2} & {\rm Vela X-1} \\
%& {\rm GX 1+4} &{\rm SXP1062}&{\rm 4U 2206+54}\\
\hline
\multicolumn{3}{c}{\hbox{Measured parameters}}\\
\hline
P^*{\hbox{(s)}} & 680 & 283 \\
%& 140 & 1062 &5560\\
P_B {\hbox{(d)}} & 41.5 & 8.96 \\
% & 1161 & \sim 300^\dag& 19\\
v_{w} {\hbox{(km/s)}} & 300 & 700 \\
%& 200 & \sim 300^\ddag& 350\\
\mu_{30}& 2.7 & 1.2 \\
%& ? & ? & 1.7\\
\dot M_{16} & 3 & 3 \\
%%& 1 & 0.6 & 0.2\\ 
%%%%%%<\dot \omega_{sd}> {\rm(rad/s)} & & & -1.5\times 10^{-12} \\
\frac{\partial \dot \omega}{\partial y} \arrowvert_{y=1}{\hbox{(rad/s}^2)} 
& 1.5\cdot10^{-12} & 1.2\cdot10^{-12} \\
%& n/a & n/a & n/a \\
% \dot\omega^*_{sd} & 0 & 0 & - 2.34 \cdot 10^{-11} & - 1.63 \cdot 10^{-11} & -9.4 \cdot 10^{-14}\\
\hline
\multicolumn{3}{c}{\hbox{Derived parameters}}\\
\hline
f(u)\zeta^{-7/11} & 0.30 & 0.32 \\
\tilde K K_1 K_3\zeta^{-3/11}& 9.1 & 7.9\\
%& & & \gtrsim 8\\
%Z& 3.7 & 2.6\\
%B_t/B_p & 0.17 & 0.22\\
%R_A{\hbox{(cm)}}& 2\cdot 10^9 & 1.4\cdot 10^9\\
%\omega^*/\omega_K(R_A)& 0.07 & 0.08\\
\frac{v_{w,min}}{\sqrt{\delta}}(1-z/Z)^{1/4}\zeta^{1/11} \hbox{(km/s)}& 540 & 800\\
%\mu_{30,min}& & &\mu_{min}'\approx4&\mu_{min}''\approx20&\mu_{min}'\approx 3.6 \\
\hline
\end{array}
$$
%$^\dag$ Estimate of the source's position in the Corbet diagram
%$^\ddag$ Estimate of typical wind velocity binary pulsars containing Be-stars.
\end{table*}

Let us apply \Eq{k1k3zeta} to two equilibrimu X-ray pulsars in which all four observable 
quantities ($\mu$, $\dot M$, $P^*$, and $\partial\omega^*/\partial \dot M$) are known:  GX 301-2
and Vela X-1 (see Table 1). The main result is that the dimensional complex 
$\tilde K K_1 K_3\zeta^{-3/11}\sim 10$ in both cases. As factors $\tilde K$, $K_1$ and
$\zeta^{-3/11}$ are of the order of one, this suggests that $K_3\sim 10$. Therefore,
the size of the 
bottom part of the boundary layer with linear angular velocity dependence on radius, where 
most of the angular momentum is transferred from the magnetosphere to the shell, $\zeta'R_A\sim 
0.1 R_A$ in both cases.

%\section{Non-equilibrium pulsars}

At $y_{cr}=(3/7)^{11/4}\approx 0.097$ the dependence $\dot \omega^*(y)$ reaches minimum 
(see Fig. \ref{f:y}).
Apparently, depending on whether $y>y_{cr}$ or $y<y_{cr}$, 
\textit{correlated changes} of $\delta \dot\omega^*$
with X-ray flux should have different signs (see Fig. \ref{f:y}). 
Indeed, for GX 1+4 in \cite{1997ApJ...481L.101C} and
\cite{2012A&A...537A..66G} a positive correlation
of the observed $\delta \dot P^*$ with $\delta \dot M$ was found using the CGRO \textit{BATSE}
and \textit{Fermi} GBM data. This means that 
there is a negative correlation between $\delta\dot \omega^*$ and $\delta\dot M$, suggesting $y<y_{cr}$ in this source. 

The application of the elaborated theory of subsonic wind accretion to non-equilibrium pulsars
is discussed in \cite{2012MNRAS.420..216S}, \cite{2013arXiv1302.0500S} and 
elsewhere in this volume \cite{KPmagbound}.

\textbf{Acknowledgements}. The authors acknowledge the Organizers of this Workshop and
RFBR grant 12-02-00186a for support. 

%
% BibTeX or Biber users please use (the style is already called in the class, ensure that the "woc.bst" style is in your local directory)
\bibliography{wind}
%\begin{thebibliography}{}
%Journal Author, Journal \textbf{Volume}, page numbers (year)
%Book Author, \textit{Book title} (Publisher, place, year) page numbers
% etc
%\end{thebibliography}

\end{document}